\documentclass{article}
\usepackage{ijcai26}

\usepackage{times}
\usepackage{soul}
\usepackage{url}
\usepackage{xspace}
\usepackage[hidelinks]{hyperref}
\usepackage[utf8]{inputenc}
\usepackage[small]{caption}
\usepackage{graphicx}
\usepackage{amsmath}
\usepackage{amsthm}
\usepackage{booktabs}
\usepackage{algorithm}
\usepackage{algorithmic}
\usepackage[switch]{lineno}
\usepackage{enumitem} 

\usepackage{latexsym}
\usepackage{xcolor}
\usepackage{natbib}
\usepackage{bbm} 
\usepackage{CJKutf8}
\usepackage{graphicx}

\definecolor{mygray}{gray}{.9}

\usepackage{array, tabularx}

\newcolumntype{Y}{>{\centering\arraybackslash}X}

\usepackage{nicematrix}
\usepackage{tikz}
\usetikzlibrary{arrows.meta, calc, positioning}
\usepackage{makecell}
\usepackage[most]{tcolorbox}
\usepackage{graphicx}
\usepackage{multirow}
\usepackage{overpic}

\newcounter{urlref}

\title{Implicit Identity Technologies for LLMs: Fingerprinting and Watermarking  \\ across Datasets, Models, and Generated Content
}

\author{
Bing Liu$^{1,2}$\and
Shunping Wang$^1$\and
Yufan Zhu$^{3,4}$\and
Xinyi Yu$^{1}$\and
Jing Huang$^{3,4}$\and \\
Linkang Du$^{1}$\and 
Hongbin Pei$^{1,5}$\thanks{Corresponding authors.}\and 
Wei Luo$^6$\\
\affiliations
$^1$School of Cyber Science and Engineering, Xi'an Jiaotong University, Xi'an, China\\
$^2$State Grid Henan Marketing Service Center, Henan, China\\
$^3$Institute of Information Engineering, Chinese Academy of Sciences, Beijing, China\\
$^4$School of Cyber Security, University of Chinese Academy of Sciences, Beijing, China\\
$^5$Shenzhen Loop Area Institute, Shenzhen, China\\
$^6$School of Information Technology, Deakin University, Geelong, Australia\\
}

\begin{document}

\maketitle

\begin{abstract}
Large language models (LLMs) require substantial investments and are increasingly deployed in high-stakes domains, making it critical to protect LLM-related assets and to trace their provenance. Identity technologies such as fingerprinting and watermarking address these needs by enabling ownership verification and attribution, and have rapidly emerged as an active research focus.
However, existing techniques lack a systematic organisation, leading to two key issues, \emph{terminological confusion} and \emph{isolated research lines}, that have hindered the development of this research field. 
To this end, we present a comprehensive review of LLM identity techniques, focusing on fingerprinting and watermarking across the LLM lifecycle, including datasets, models, and generated content. 
We make three primary contributions. \emph{First}, we introduce \emph{implicit identity} (Implicit-ID for short) as a unifying abstraction and distinguish fingerprinting from watermarking. 
\emph{Second}, we propose a lifecycle-based taxonomy that organises techniques by asset type and verification role, aligning each with asset protection or provenance.
\emph{Third}, we establish an evaluation framework around three objectives---identifiability, robustness, and deployability. 
Together, these contributions structure the landscape of LLM identity techniques, clarify terminology, and highlight directions toward secure deployment.

\end{abstract}

\section{Introduction}
Large language models (LLMs) have revolutionised a wide range of fields.
Their development requires substantial investments in data, computation, and expertise; for example, training GPT-4 is reported to cost over \$100 million\footnote{\url{https://en.wikipedia.org/wiki/GPT-4}}
.

Consequently, \emph{asset protection} of LLMs becomes critical given such high costs, as unauthorised use or leakage can result in significant losses.
In parallel, the misuse and malicious exploitation of LLMs has emerged as another key challenge; for instance, LLM-generated content has been used by malicious actors to enable automated fraud systems \citep{lin2024malla}.
Accordingly, \emph{provenance} of LLMs is critical for tracing the origins of such activities, thereby enabling accountability and governance of misuse and malicious exploitation.




From a technological perspective, identity mechanisms for LLMs are fundamental to enabling both asset protection and provenance.
A growing body of research has explored LLM identity techniques, such as fingerprinting for model copyright protection~\citep{zhang2024reef} and watermarking for generated content provenance~\citep{kirchenbauer2023watermark}.
However, as the field remains at an early stage, existing technologies lack a systematic organisation, resulting in two key issues:
(1) \textbf{Terminological confusion}. In particular, watermarking and fingerprinting, two primary identity technologies, are not clearly distinguished in the literature.
For instance, the trigger-based specific-response mechanisms are termed “instructional fingerprinting” in~\citep{xu2024instructional}, whereas they are referred to as “backdoor watermarking” in~\citep{mo2025tibw}.
(2) \textbf{Isolated research lines}. 
Existing identity studies for LLMs have largely evolved in isolation within specific scenarios, including dataset identity~\citep{carlini2021extracting,maini2021dataset}, model identity  ~\citep{zeng2024huref,zhang2024reef,xu2024instructional}, and generated content identity  ~\citep{kirchenbauer2023watermark,hou2024semstamp,hu2023unbiased}. 
As a result, the community lacks a comprehensive and unified understanding of LLM identity.
Prior watermarking surveys \citep{liu2024survey,wang2025building,liang2026watermarking} naturally do not cover fingerprinting, and even recent surveys covering both  \citep{xu2025copyright,ye2025securing,wang2025building,liang2026watermarking} fail to adequately address these two issues.

The two aforementioned issues have, to some extent, hindered the development of this research field.
To address these challenges, this survey clarifies key concepts and proposes a verification-semantic taxonomy to systematically organise existing studies on LLM identity, including fingerprinting and watermarking.
Our goal is to clarify the landscape of this nascent research field and to outline research objectives and major roadmaps to guide future advances.
Therefore, this survey makes the following three contributions.


\textbf{Our first contribution is a unification of key concepts in this field}.
We propose a three-layer conceptual framework to organise existing methods.
(I) At the top level, we introduce a unifying abstraction, \emph{Implicit Identity} (Implicit-ID for short), which refers to specific characteristics of an LLM that are not directly observable but can be verified and used to uniquely identify the model and distinguish it from others. Such implicit identity provides a unifying abstraction for both fingerprinting and watermarking techniques.
(II) At the middle level, we differentiate between two main realisations of implicit-ID, fingerprinting and watermarking. LLM fingerprinting is defined as a \emph{non-intrusive} implicit-ID, which is constructed by intrinsic characteristics of an LLM; LLM watermarking is an \emph{intrusive} implicit-ID, in which identifiable signals are deliberately embedded into the model, data, or generated content through external intervention.
(III) At the bottom level, we summarise existing methods by evidence source and verification pathway, yielding a compact design-space view of implicit-ID technologies.


\textbf{Our second contribution is a generative taxonomy across assets and lifecycle stages}.
We introduce a taxonomy that organises identity technologies \emph{by verification semantics}, i.e., similarity-based attribution and keyed verification, across datasets, models, and generated content throughout the LLM lifecycle. Beyond providing comprehensive coverage, the taxonomy reveals \emph{cross-asset equivalences}.
For instance, the shared trigger-response logic underlying dataset watermarking and fine-tuning-based model watermarking, and the shared attribution logic underlying behavioural model fingerprints and output-only fingerprints. 
This perspective enables \emph{cross-level reasoning}: techniques designed for one asset level can inform or transfer to another, helping the community move beyond isolated research lines and supporting reasoning about multi-level identity designs for LLMs.

\textbf{Our third contribution is the establishment of technological objectives}.
Motivated by the two core application tasks of Implicit-ID, asset protection and provenance, we identify three key objectives for the technologies: \emph{identifiability}, \emph{robustness}, and \emph{deployability}.
Specifically, an Implicit-ID should exhibit high identifiability to reliably distinguish different LLMs, maintain robustness under legitimate post-deployment transformations and adversarial manipulation, and demonstrate practical deployability under realistic settings. Accordingly, we propose a common evaluation template organised by transformation regimes, enabling systematic and comparable assessment of existing methods.

These contributions show that LLM fingerprinting and watermarking can be consistently viewed under a shared verification-semantic lens, rather than as isolated, mechanism-specific techniques, as organised in the prior surveys \citep{xu2025copyright,ye2025securing}.

\paragraph{What this survey enables.}
By unifying identity technologies under Implicit-ID and organizing them by verification semantics rather than mechanism, this survey enables: (i) principled analysis and comparison between intrusive and non-intrusive methods under shared verification semantics; (ii) transfer of techniques and evaluation stressors across dataset, model, and content identity; and (iii) systematic reasoning about trade-offs between legal defensibility, robustness to post-deployment transformations, and practical deployability.

The remainder is organised as follows: Section~2 formalises the tasks and notation, Section~3 presents the lifecycle-based taxonomy, Section~4 defines the evaluation metrics, and Section~5 discusses challenges and directions.



\section{Main Tasks in Implicit-ID Technologies}
\label{section:Main Tasks in Identity Technologies}

In this section, we formalise two primary application tasks of Implicit-ID technologies: \textbf{Asset Protection}, which aims to safeguard the security and value of LLM-related assets, and \textbf{Provenance}, which supports accountability, traceability, and regulatory compliance within the LLM ecosystem.

\paragraph{Task formulation.} We view identity verification over LLM assets as the foundation of both tasks. In general, identity verification can be formulated as a decision process over observable evidence extracted from an asset, where the verifier determines whether the asset should be attributed to a claimed owner, creator, or source. This process must control both false matches, where an unrelated asset is incorrectly accepted, and false misses, where a legitimate or derived asset is incorrectly rejected, under realistic access constraints. This formulation motivates our later focus on identifiability, robustness, and deployability.
In asset protection, identity verification establishes legitimate ownership and prevents unauthorised use. In provenance, it identifies asset creators or contributors, thereby enabling accountability, traceability, and compliance.

\paragraph{Unified view and notation.}
Let $A \in \{\mathcal{D}, \mathcal{M}, \mathcal{Y}\}$ denote an LLM asset subject to protection or provenance verification, where $\mathcal{D}$ represents training data, $\mathcal{M}$ represents model artefacts such as weights and internal states, and $\mathcal{Y}$ represents generated content.
$A$ verifier accesses the asset through an interface $\mathsf{O}_A$, which captures practical verification settings: white-box access, where model weights or full asset details are available; grey-box access, where only limited internal information is exposed; and black-box access, where verification is restricted to queries.
After release or deployment, the asset may undergo a transformation $\tau \in \mathcal{T}$, which yields $A'=\tau(A)$. Examples include dataset filtering, model fine-tuning, quantisation, distillation, merging, and output rewriting.
We evaluate an Implicit-ID technology by whether reliable verification can be maintained under $\tau$ of interest.
We distinguish two transformation regimes that recur across all asset types.
\emph{Owner/benign transformations} $\tau \in \mathcal{T}_{\text{own}}$ correspond to legitimate post-release adaptations, such as filtering, fine-tuning, quantisation, distillation, prompt or policy modification, and paraphrasing. These transformations primarily evaluate whether an identity signal remains stable under normal lifecycle evolution.
\emph{Adversarial transformations} $\tau \in \mathcal{T}_{\text{adv}}$ represent adaptive attempts to remove, forge, or evade identity signals, including watermark removal through targeted training, paraphrase-based sanitisation, trigger suppression, and mimicry prompting.


\subsection{Asset Protection Task}
For LLM-related asset protection, Implicit-ID resides in or is embedded within assets and serves as verifiable evidence of ownership, thereby detecting and countering unauthorised use and redistribution, such as the well-known 2024 \textit{Miqu} model leak on Hugging Face\footnote{\url{https://huggingface.co/miqudev/miqu-1-70b}}
.
Generated content asset protection is also crucial for preventing plagiarism.
The U.S. Copyright Office’s rulings on creative arrangement in AI-assisted works establish an important legal foundation\footnote{\url{https://www.copyright.gov/docs/zarya-of-the-dawn.pdf}}

Asset protection can be achieved through proactive verification, i.e., watermarking. Specifically, the asset owner embeds a keyed identity signal into an asset (or its generation process) using an embedding function $E_k$, producing a protected asset $\widetilde{A}=E_k(A)$. After the asset undergoes a transformation $A'=\tau(\widetilde{A})$, ownership is verified through a keyed detector $V_k(\mathsf{O}_{A'})$.
Asset protection can also be achieved through similarity-based attribution, i.e., fingerprinting, where no keyed mark is embedded. Instead, the verifier extracts inherent distinguishing evidence $X(\mathsf{O}_{A'})$ from the target asset and determines ownership by comparing it with reference evidence extracted from owned assets, through a similarity-based detector $V_x(\mathsf{O}_{A}, \mathsf{O}_{A'})$.



\vspace{-1mm}
\subsection{Provenance Task}
For provenance, Implicit-ID acts as the foundation for tracing LLM-related asset lineage and origin, reinforcing digital trust and accountability.
Dataset provenance enables retroactive auditing through Implicit-ID techniques such as dataset fingerprinting, allowing stakeholders to detect latent “data footprints” \citep{sadhukhan2025footprints} and verify licensing compliance without requiring pre-marked data.
Model provenance focuses on identifying rebranded or derivative models and tracing them back to their foundational counterparts, supporting regulatory requirements for a transparent chain of custody.
Generated content provenance usually serves as a societal safeguard against misinformation and synthetic fraud. 
Many jurisdictions have established mandatory labelling requirements for generated content, including China\footnote{\url{https://www.mps.gov.cn/n6557558/c8797736/content.html}}
, the EU\footnote{\url{https://artificialintelligenceact.eu/article/50/}}
, and the United States\footnote{\url{https://leginfo.legislature.ca.gov/faces/billNavClient.xhtml?bill_id=202320240SB942}}
, thereby reinforcing digital trust through provenance.


In practice, provenance can be established either through \emph{proactive provenance binding}, e.g., watermarking at creation, or through \emph{retroactive provenance inference} similarity-based attribution when no prior mark exists.
Provenance binding is a proactive realisation of provenance. In this mode, the asset owner embeds or attaches verifiable evidence using $E_k$, such as a keyed watermark, either at creation time or during service \citep{dathathri2024scalable}. 
The resulting asset carries an identity signal that can later be verified from a transformed asset $A'$ using a keyed detector $V_k(\mathsf{O}_{A'})$. This mode provides strong, deliberate provenance evidence, but typically requires prior instrumentation of the asset or serving pipeline.
Provenance inference is a retroactive realisation of provenance. When no prior mark is assumed, the verifier extracts inherent evidence $X(\mathsf{O}_{A'})$ from the observed asset and attributes it by comparing this evidence with reference evidence from candidate source assets through a matcher $V_x(\mathsf{O}_{A}, \mathsf{O}_{A'})$. This mode enables post hoc auditing, and its conclusions are generally probabilistic rather than cryptographically guaranteed.

\section{Taxonomy across LLM Lifecycle and Verification Semantics}
\label{section:Classification of Identity Technologies Across the LLM Lifecycle}

We first introduce the taxonomy dimensions that structure the landscape of Implicit-ID technologies, including the LLM lifecycle and verification semantics, and then systematically survey existing techniques within this unified framework.


\begin{figure}[t]
    \centering
    \includegraphics[width=\columnwidth]{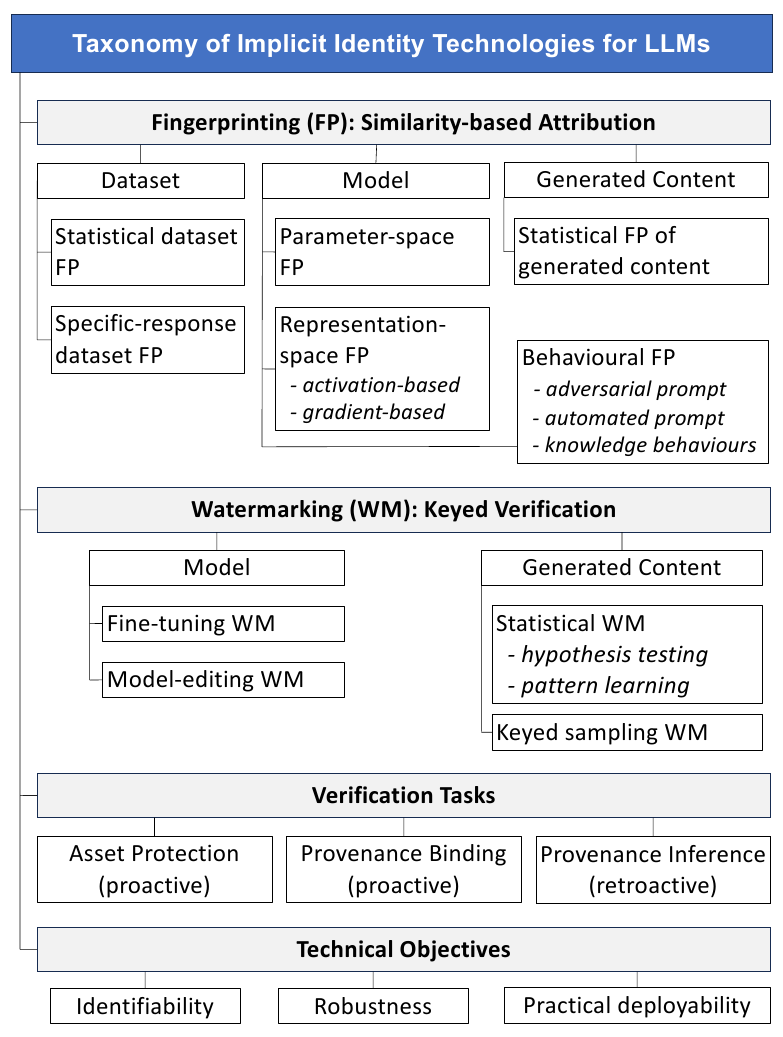} 
    \caption{Taxonomy of Implicit-ID technologies for LLMs.}
    \label{fig:Taxonomy of Identity Technologies}
\end{figure}

\subsection{Taxonomy Dimensions}

Our taxonomy organises Implicit-ID technologies along two complementary dimensions: a \emph{verification dimension} defined by verification semantics, and a \emph{lifecycle dimension} defined by the asset stage at which identity is established or verified.

Along the lifecycle dimension, we consider three asset stages throughout the LLM lifecycle: Dataset $\mathcal{D}$, Model $\mathcal{M}$, and Generated content $\mathcal{Y}$. 
Asset protection and provenance tasks recur as cross-cutting objectives across all stages.

Along the verification dimension, we distinguish two categories of verification semantics underlying both asset protection and provenance tasks. 
The first category is \textbf{similarity-based attribution}, implemented via a matcher $V_x(\cdot,\cdot)$, which corresponds to \textbf{non-intrusive fingerprinting}.
The second category is \textbf{keyed verification} $V_k(\cdot)$, implemented through a keyed mark and a corresponding detector, which corresponds to \textbf{intrusive watermarking}.
Consequently, both asset protection and provenance can be achieved either through keyed verification or through similarity-based attribution.

Specifically, across assets $A \in {\mathcal{D}, \mathcal{M}, \mathcal{Y}}$ and corresponding access interfaces $\mathsf{O}_A$, we summarize existing methods in terms of:
(i) \textbf{Evidence source} and required \textbf{Access conditions};
(ii) \textbf{Verification semantics} (keyed or similarity-based); and
(iii) \textbf{Dominant transformation stressors} $\tau$ under which verification must remain reliable.
The proposed taxonomy is \emph{generative}: it reveals cross-asset commonalities in verification logic, suggesting how techniques, attack models, and evaluation stressors may transfer across datasets, models, and generated content. 
This perspective enables cross-level reasoning, where techniques developed for one asset stage can inform the design of methods at other stages.

\subsection{Implicit-ID of Datasets}
\label{subsection:Identity Technologies for Datasets}

Implicit-ID of datasets concerns verifying whether a suspect model $\mathcal{M}'$ was trained on a target dataset $\mathcal{D}$ under post-training transformations $\tau$, given access through an observation interface, most commonly a black-box API.
Its primary task is \emph{provenance inference}: providing post-hoc evidence of data usage when training provenance is opaque, thereby supporting auditing and compliance; by establishing data ownership, it also indirectly contributes to asset protection.

Although dataset watermarking has been studied in traditional machine learning, particularly for image datasets~\citep{du2025sok}, emerging LLM-oriented approaches embed in-distribution, text-preserving signals into documents, such as invisible Unicode canaries or cue–reply structures, and later verify their imprint through black-box prompting~\citep{li2025data,naser2025auditing}. 
Since these methods typically rely on special triggers that induce recognisable model behaviour after training, we categorise them as fine-tuning  model watermarking and discuss them in Section~3.3.
This subsection focuses on dataset fingerprinting for provenance inference, where the verifier does not assume a pre-embedded mark.

\vspace{0.3em}
\noindent\fbox{%
  \begin{minipage}{\dimexpr\linewidth-2\fboxsep-2\fboxrule}
  \setlength{\parskip}{0pt}
  \setlength{\baselineskip}{0.95\baselineskip}
  
  \textbf{Dataset fingerprint}
  \begin{itemize}[leftmargin=*, topsep=1pt, itemsep=0.5pt, parsep=0pt]
    \item \textbf{Evidence:} loss/likelihood statistics.
    \item \textbf{Access:}  black-box or grey-box.
    \item \textbf{Verification:}  statistical memorisation signal detection, targeted behavioural signature probing.
    \item \textbf{Stressors:}  shift, fine-tuning, distillation.
  \end{itemize}
  \end{minipage}
}
\vspace{0.1em}

Dataset fingerprinting can be broadly divided into two paradigms: \textbf{statistical fingerprinting}, which relies on aggregate memorization signals, and \textbf{specific-response fingerprinting}, which examines targeted behavioural signatures.

\subsubsection{Statistical Dataset Fingerprinting} 
Statistical dataset fingerprinting provides non-keyed evidence that $\mathcal{M}'$ has been exposed to $\mathcal{D}$ by identifying dataset-specific \emph{memorization signals} in model statistics,  including loss or perplexity gaps, likelihood anomalies, and aggregate membership scores, while leaving $\mathcal{D}$ unmodified.

Representative lines of work show that LLMs can exhibit detectable membership effects at the sequence or aggregate level, enabling post-hoc auditing and ownership resolution 
\citep{carlini2021extracting,mattern2023membership,shi2023detecting,maini2024llm}. In practice, these signals are most informative when evaluated under confound controls (e.g., semantically similar corpora) and robustness stressors $\tau$ (fine-tuning, distillation, and distribution shift), which can dilute or mask membership traces \citep{huang2024general}.

\subsubsection{Specific-Response Dataset Fingerprinting}

Specific-response fingerprinting constructs non-keyed evidence by probing $\mathcal{M}'$ with a curated set of membership-sensitive queries and analysing \emph{targeted behavioural signatures} rather than global statistics. Typical signals include confidence/margin patterns on high-influence examples or specialised probe prompts that elicit distinctive responses linked to $\mathcal{D}$ \citep{maini2021dataset,liu2022your,naser2025auditing}.

Compared with purely statistical approaches, this paradigm can offer stronger diagnostic power for ownership resolution, but is more sensitive to probe design and can be confounded by \emph{independent similarity} (training on a dataset $\mathcal{D}' \approx \mathcal{D}$). It is also vulnerable to post-training transformations $\tau$ (e.g., instruction tuning) and adaptive smoothing that reduce membership contrast.

\subsection{Implicit-ID of Models}
\label{subsection:Identity Technologies for Models}

Implicit-ID of models concerns determining whether a suspect model $\mathcal{M}'$ is derived from an owner model $\mathcal{M}$ under transformations $\tau \in \mathcal{T}$ and an observation interface, which may provide white-, grey-, or black-box access. Such identity verification supports both asset protection and provenance.
Here, we focus on model-level identity, covering \emph{fingerprinting as similarity-based attribution} and \emph{watermarking as keyed ownership verification}.
We defer output provenance techniques to Section~\ref{subsection:Identity Technologies for Outputs}.

\subsubsection{Model Fingerprinting}


Model fingerprinting performs non-keyed, similarity-based attribution by matching intrinsic model evidence, such as parameters, representations, or elicited behaviors.

Existing works form fingerprints by leveraging model parameters, internal representations, and externally observable behaviours.
 Accordingly, existing methods fall into three categories: \textbf{parameter-space fingerprints}, which require white-box access; \textbf{representation-space fingerprints}, which typically require white- or grey-box access, with occasional query-only approximations; and \textbf{behavioural fingerprints}, which rely on curated elicitation prompts and are compatible with black-box access.



\vspace{0.3em}
\noindent\fbox{%
  \begin{minipage}{\dimexpr\linewidth-2\fboxsep-2\fboxrule}
  \setlength{\parskip}{0pt}
  \setlength{\baselineskip}{0.95\baselineskip}
  
  \textbf{Model parameter-space fingerprint}
  \begin{itemize}[leftmargin=*, topsep=1pt, itemsep=0.5pt, parsep=0pt]
    \item \textbf{Evidence:} weights/parameters statistics/invariants.
    \item \textbf{Access:} white-box.
    \item \textbf{Verification:} similarity-based attribution.
    \item \textbf{Stressors:} weight permutation/reordering, mild fine-tuning, typically weak under strong $\tau$
    (e.g., distillation or merging) unless explicitly handled.
  \end{itemize}
  \end{minipage}
}
\vspace{0.1em}

Model parameter-space fingerprinting identifies a model by extracting descriptors directly from weights, then attributing ownership via similarity in a weight-derived feature space.
One representative design is to build invariants that remain stable under admissible weight reordering/permutation (e.g., HuRef~\citep{zeng2024huref}), enabling lineage/copyright inference without embedding any trigger.
Another design~\citep{yoon2025intrinsic} uses layerwise attention-projection statistics (Q/K/V/out) and compares correlations across statistical sequences to infer lineage (Intrinsic Fingerprint of LLMs).
A third line of work~\citep{zeng2025awm} aligns informative weight blocks (e.g., embeddings and Q/K projections) and computes similarity after alignment (AWM), reported as discriminative even after large-scale training.
In exchange for strong identifiability signals, parameter-space approaches typically require white-box access and may be costly or less practical for ultra-large models.

\vspace{0.2em}
\noindent\fbox{%
  \begin{minipage}{\dimexpr\linewidth-2\fboxsep-2\fboxrule}
  \setlength{\parskip}{0pt}
  \setlength{\baselineskip}{0.95\baselineskip}
  \textbf{Model representation-space fingerprint}
  \begin{itemize}[leftmargin=*, topsep=1pt, itemsep=0.5pt, parsep=0pt]
    \item \textbf{Evidence:} hidden activations/gradients 
    under probing.
    \item \textbf{Access:} white-box, grey-box, black-box occasionally.
    \item \textbf{Verification:} similarity-based attribution.
    \item \textbf{Stressors:} probe design and distribution shift, fine-tuning, merging, distillation.
  \end{itemize}
  \end{minipage}
}
\vspace{0.1em}

Model representation-space fingerprinting attributes a suspect model by comparing its internal responses to predefined probing inputs. These responses include activations, gradients, or sensitivity signals, which are converted into descriptors and matched against those of reference models.


Existing approaches can be organised into two canonical designs.
The first is \emph{activation-based fingerprints}, which identify models by comparing intermediate activations induced by probing inputs:
\textsc{DeepJudge}~\citep{chen2022copy} evaluates similarity via multi-level distance metrics on intermediate activations; 
\textsc{EasyDetector}~\citep{zhang2024easydetector} trains probing classifiers on a victim model’s activations and tests whether separability patterns transfer to a suspect model; 
and REEF~\citep{zhang2024reef} compares activation representations using geometry-preserving similarity (CKA). 
The second is \emph{gradient/sensitivity-based  fingerprints}, which use how the model responds to small perturbations: 
\textsc{TensorGuard}~\citep{wu2025gradient} extracts fingerprints from statistics of parameter gradients induced by controlled perturbations in selected layers; 
and \textsc{ZeroPrint}~\citep{shao2025reading}  derives global Jacobian-based fingerprints by aggregating output differences under probing inputs, enabling identification without access to internal states. In general, representation-space fingerprints can be robust to parameter modifications. However, most existing methods require white-box access and may struggle to distinguish closely related LLMs with similar architectures.
\vspace{1mm}

Model behaviour fingerprinting attributes a suspect model by eliciting stable and discriminative response patterns from a compact set of curated prompts, followed by non-keyed matching over the observed outputs through similarity scoring or learned linkage.
A representative instantiation is LLMMAP~\citep{pasquini2024llmmap}, which actively issues a small set of queries to fingerprint LLM-integrated systems, such as RAG pipelines, and identify specific model versions from subtle behavioural nuances.

\vspace{0.2em}
\noindent\fbox{%
  \begin{minipage}{\dimexpr\linewidth-2\fboxsep-2\fboxrule}
  \setlength{\parskip}{0pt}
  \setlength{\baselineskip}{0.95\baselineskip}
  \textbf{Model behaviour fingerprint} 
  \begin{itemize}[leftmargin=*, topsep=1pt, itemsep=0.5pt, parsep=0pt]
      \item 
  \textbf{Evidence:} input-output response patterns under curated elicitation prompts, such as fixed-string responses, response distributions, and reasoning traces.
  \item \textbf{Access:} black-box. 
  \item \textbf{Verification:} similarity-based attribution, learned matching over responses.
  \item \textbf{Stressors:} decoding randomness (temperature/top-$p$), refusal/safety policies, prompt filtering/sanitisation, system-prompt differences, context-window effects, adaptive prompt interventions, model drift.
    \end{itemize}
  \end{minipage}
}
\vspace{0.1em}

Behavioural fingerprinting can be grouped into three technical paradigms.
The first is \emph{adversarial/optimised prompting}, which aims to induce deterministic fixed-string or high-entropy signatures that separate the target model from non-homologous models. TRAP~\citep{gubri2024trap} repurposes adversarial suffixes to induce a predefined answer from the target model, while other models produce incoherent or divergent outputs.
\textsc{ProFLingo}~\citep{jin2024proflingo} generates prompts that trigger distinctive response distributions to link original models and their derivatives.
The second is \emph{automated prompt generation}, which reduces reliance on manually crafted fingerprints.
Hide-and-Seek~\citep{iourovitski2024hide}  uses evolutionary learning with an auditor LLM to discover salient input–output pairs exposing a model family’s semantic manifold; 
RAP-SM~\citep{xu2025rap} further leverages shadow models to improve fingerprint transferability across model series and maintain robustness under adaptive transformations such as model merging and fine-tuning.
The third is based on \emph{high-level cognitive and knowledge-level behaviour}, where reasoning and retrieval idiosyncrasies serve as fingerprints. 
\textsc{CoTSRF}\citep{Ren2025CoTSRFUC} models distinctive chain-of-thought trajectories elicited by crafted logic prompts, whereas \textsc{DuFFin}\citep{Yan2025DuFFinAD} combines trigger patterns with knowledge-level fingerprints that capture how models retrieve and articulate domain-specific facts.
Additionally, some behavioural identifiers exploit architecture-tied effects. For example, \textsc{RouteMark}~\citep{he2025routemark} attributes identity in MoE-based merged models through distinctive expert-routing logic. Although such evidence is observed behaviourally, it reflects internal architectural characteristics and therefore lies at the boundary between behaviour and representation-space fingerprinting.

Taken together, these methods show that input-conditioned responses—whether triggered by adversarial suffixes, reasoning paths, or routing preferences—constitute a powerful and diverse toolkit for active model fingerprinting.

\subsubsection{Model Watermarking}

Model watermarking refers to keyed ownership verification: a class of intrusive identity technologies that embed identity signals into a model before release, allowing ownership to be verified later by probing for those signals.
Most existing methods follow a \emph{specific-response} paradigm. The model is intentionally modified to produce distinctive responses to designated secret triggers while preserving normal behavior on benign inputs. These trigger–response behaviors then serve as explicit identifiers for ownership verification.

Such methods are often referred to as ``fingerprinting" or ``backdoor fingerprinting," but this terminology can be misleading.
Unlike fingerprints that arise intrinsically from architecture or training data, watermarks are deliberately embedded identity signals. 
We therefore categorise these specific-response methods as model watermarking.


From the perspective of watermark embedding mechanisms, existing model watermarking methods can be broadly divided into two categories: \emph{fine-tuning-based watermarking} and \emph{model-editing-based watermarking}. The key distinction lies in whether the watermark is implanted through additional training or by directly editing model parameters.

\vspace{0.2em}
\noindent\fbox{%
  \begin{minipage}{\dimexpr\linewidth-2\fboxsep-2\fboxrule}
  \setlength{\parskip}{0pt}
  \setlength{\baselineskip}{0.95\baselineskip}
  \textbf{Model fine-tuning watermark}
  \begin{itemize}[leftmargin=*, topsep=1pt, itemsep=0.5pt, parsep=0pt]
    \item \textbf{Evidence:} trigger-response behaviour embedded.
    \item \textbf{Access:} black-box.
    \item \textbf{Verification:} keyed verification.
    \item \textbf{Stressors:} suppression/filtering, prompt interventions, extraction, fine-tuning, merging, distillation.
  \end{itemize}
  \end{minipage}
}
\vspace{0.1em}

Model watermarking via fine-tuning represents an early and widely used paradigm for embedding trigger–response behaviours into LLMs. In this setting, the watermark is injected by fine-tuning the model on data containing secret trigger–response pairs, thereby inducing the model to associate predefined triggers with designated responses.

Instructional fingerprinting~\citep{xu2024instructional} demonstrates that instruction-style supervised fine-tuning can implant watermark behaviours without requiring access to model internals. 
Subsequent work extends the approach via stronger trigger-response binding, as in Chain\&Hash~\citep{russinovich2024hey}; improved deployment efficiency, as in FP-vec~\citep{xu2024fp}; greater scalability via in-distribution trigger–response pairs, as in Perinucleus Fingerprinting~\citep{nasery2025scalable}; and enhanced robustness to further fine-tuning by anchoring watermark behaviours in lower-level representations, as in TIBW~\citep{mo2025tibw}.

The primary stressors for fine-tuning-based watermarks include suppression or filtering, prompt-level interventions, etc. 
Because many model watermarks are implemented through trigger–response behaviours, their robustness should also be evaluated against adjacent backdoor-repair and unlearning techniques. For example, class-wise trigger recovery and unlearning suggest that embedded trigger behaviours can be actively discovered and weakened~\citep{hou2025fixguard}.

\vspace{0.2em}
\noindent\fbox{%
  \begin{minipage}{\dimexpr\linewidth-2\fboxsep-2\fboxrule}
  \setlength{\parskip}{0pt}
  \setlength{\baselineskip}{0.95\baselineskip}
  \textbf{Model-editing watermark}
  \begin{itemize}[leftmargin=*, topsep=1pt, itemsep=0.5pt, parsep=0pt]
    \item \textbf{Evidence:} keyed behaviours inserted.
    \item \textbf{Access:} black-box.
    \item \textbf{Verification:} keyed verification.
    \item \textbf{Stressors:} subsequent edits/unlearning, fine-tuning, merging, distillation.
  \end{itemize}
  \end{minipage}
}
\vspace{0.05em}

Model-editing watermarking inserts keyed behaviours through localized parameter edits, enabling low-cost watermark implantation with limited impact on model utility. 
Representative designs either reinforce trigger–target associations, as in FPEdit~\citep{wang2025fpedit}, or embed more natural keyed QA behaviours, as in EditMark~\citep{li2025editmark}. 
We evaluate the robustness of model watermarks under these stressors and defer a detailed discussion of the corresponding attack surface and evaluation protocols to Sec.~\ref{sec:metrics}.



\vspace{-1mm}
\subsection{Implicit-ID of Generated Content}
\label{subsection:Identity Technologies for Outputs}

Implicit-ID of generated content concerns determining whether a given output $\mathcal{Y}$ originates from a specific model $\mathcal{M}$, solely based on the output text under post-generation transformations $\tau \in \mathcal{T}$ and black-box access.

We distinguish two verification semantics: \textbf{generated content fingerprinting}, which exploits inherent idiosyncrasies in model responses to perform provenance inference; and \textbf{generated content watermarking}, which intentionally embeds a detectable signal during generation to support provenance binding and proactive asset protection.





\subsubsection{Generated Content Fingerprinting}

\vspace{0.1em}
\noindent\fbox{%
  \begin{minipage}{\dimexpr\linewidth-2\fboxsep-2\fboxrule}
  \setlength{\parskip}{0pt}
  \setlength{\baselineskip}{0.95\baselineskip}
  \textbf{Generated content fingerprint}
\begin{itemize}[leftmargin=*, topsep=1pt, itemsep=0.5pt, parsep=0pt]
   \item \textbf{Evidence:} distributional/stylistic biases in outputs.
  \item \textbf{Access:} black-box.
  \item \textbf{Verification:} similarity-based attribution.
  \item \textbf{Stressors:} rewriting, decoding changes, confounding.
\end{itemize}
\end{minipage}}
\vspace{0.1em}

Generated content fingerprinting attributes a source model from \emph{outputs alone} by extracting
\emph{intrinsic} and \emph{unintended} regularities in the generated text, and performing non-keyed attribution, e.g., similarity scoring or classifier-based matching without modifying the model or decoding procedure.

The main technology is statistical fingerprinting of generated content, which treats stable deviations in an LLM’s output distribution, whether lexical, syntactic, or embedding-level, as latent signatures for attribution through statistical modeling or classification.
Recent studies report strong discriminability at the \emph{model-family} level. \citet{sun2025idiosyncrasies} show that word-level distributional cues enable embedding-based classifiers to distinguish major model families, such as GPT, Claude, and Gemini, with robustness under rewriting and translation. \citet{suzuki2025natural} further suggest that identifiable ``natural fingerprints’’ can emerge even among models trained on identical data, owing to randomness in optimization. From a detection standpoint, \citet{antoun2023text} demonstrate cross-model attribution that generalizes across sources and model sizes, while \citet{bitton2025detecting} improve reliability through classifier ensembles that preserve precision under style-mimic prompts.
Because these signals are distributional, attribution is sensitive to prompt/system context and decoding settings. It is most defensible when framed as probabilistic evidence rather than cryptographic proof.

\subsubsection{Generated Content Watermarking}

Generated content watermarking is an intrusive provenance-binding mechanism that intentionally embeds verifiable identity signals into the text generation process, enabling provenance to be verified directly, without requiring access to model internals.
We organize existing methods according to how identity signals are embedded. The first is \emph{statistical watermarking}, which introduces detectable distributional biases into generated text, either through inference-time manipulation of token probabilities or through training-time internalization of watermark behaviors. The second is \emph{keyed sampling watermarking}, which encodes identity within keyed sampling trajectories, enabling provenance verification with little or no perceptible distortion to the generated content.

The statistical watermarking embeds a \emph{verifiable distributional deviation} into generated text such that provenance can be detected from outputs alone. Whether injected via inference-time logit intervention or internalised during training, the watermark manifests as a persistent statistical shift detectable by a verifier. Based on the verification mechanism, this paradigm admits two dominant technical paths: \emph{statistical hypothesis testing} and \emph{statistical pattern learning}.

\vspace{0.1em}
\noindent\fbox{%
  \begin{minipage}{\dimexpr\linewidth-2\fboxsep-2\fboxrule}
  \setlength{\parskip}{0pt}
  \setlength{\baselineskip}{0.95\baselineskip}
  \textbf{Statistical watermark of generated content}
\begin{itemize}[leftmargin=*, topsep=1pt, itemsep=0.5pt, parsep=0pt]
    \item \textbf{Evidence:} verifiable distributional deviation.
    \item \textbf{Access:} black-box.
    \item \textbf{Verification:} keyed verification.
    \item \textbf{Stressors:} rewrite\allowbreak/\allowbreak translate\allowbreak/\allowbreak summarise\allowbreak/\allowbreak character-level perturbations, decoding changes, adaptive removal.
\end{itemize}
\end{minipage}}
\vspace{0.1em}


The line of statistical hypothesis testing formulates provenance verification as a statistical hypothesis test, where the null hypothesis assumes natural, unwatermarked text. \citep{kirchenbauer2023watermark} introduce the canonical \emph{red–green list} scheme, which biases token selection toward a pseudo-random subset of the vocabulary and verifies provenance through Z-score testing of token frequencies. \citep{dathathri2024scalable} refine this framework in \textsc{SynthID-Text} by using tournament sampling to preserve generation diversity while maintaining high detection specificity.
To mitigate the fragility of token-level statistics under paraphrasing, subsequent work shifts verification into semantic space. 
~\citep{hou2024semstamp} propose \textsc{SemStamp}, which partitions embedding space using locality-sensitive hashing and treats region occupancy as a statistical signature; 
\textsc{K-SemStamp}~\citep{hou2024k} further aligns partitions with semantic structure via clustering, improving robustness to linguistic transformations. Across these methods, watermark verification remains an \emph{external, rule-based} statistical test.


The line of statistical pattern learning treats the watermark as a \emph{latent statistical pattern} that can be internalized by the model and recognized by learned discriminators. This approach is particularly relevant in open-source settings, where inference-time logit biasing can be bypassed.
~\citep{gu2023learnability} demonstrate watermark distillation, showing that student models can reproduce a teacher’s statistical watermark without explicit decoding constraints. 
~\citep{xu2024learning} further improves detectability via reinforcement learning co-training, coupling generation with a trainable detector to shape output trajectories that maximise classification confidence while preserving fluency. This notion of internalised identity is extended to open-source security in \textsc{Mark-Your-LLM}~\citep{xu2025mark}, 
which embeds statistical behaviours via backdoor-based internalisation and verifies provenance using a paired classifier.

Keyed sampling watermarking encodes provenance through \emph{secrecy in the sampling process}. 
Keyed sampling defines the watermark through secret sampling rules, typically keyed pseudo-randomness governing token selection, rather than through observable deviations in the output distribution. Verification then tests whether a generated sequence is consistent with the secret key and the model’s logits.
A central advantage of this paradigm is its \emph{zero-distortion} property: the marginal distribution of watermarked text is theoretically identical to that of the original model. 
~\citep{christ2024undetectable} formalises this notion via \emph{undetectable watermarks} based on cryptographic one-way functions, showing that outputs are computationally indistinguishable from unwatermarked text to any adversary lacking the key. 
~\citep{hu2023unbiased} further propose an unbiased watermarking framework based on reparameterized sampling, ensuring that expected token frequencies match the model’s softmax distribution.

\vspace{0.1em}
\noindent\fbox{%
  \begin{minipage}{\dimexpr\linewidth-2\fboxsep-2\fboxrule}
  \setlength{\parskip}{0pt}
  \setlength{\baselineskip}{0.95\baselineskip}
  \textbf{Keyed sampling  watermark of generated content}
\begin{itemize}[leftmargin=*, topsep=1pt, itemsep=0.5pt, parsep=0pt]
    \item \textbf{Evidence:} keyed sampling trace.
    \item \textbf{Access:} black-box.
    \item \textbf{Verification:} keyed verification.
    \item \textbf{Stressors:} rewrite\allowbreak/\allowbreak translate\allowbreak/\allowbreak summarise\allowbreak/\allowbreak character-level perturbations, decoding changes, adaptive removal.
\end{itemize}
\end{minipage}}
\vspace{0.1em}

In keyed sampling watermarking, verification reduces to re-evaluating whether the observed token sequence adheres to the secret sampling trajectory induced by the key and the model probabilities. By exploiting the secrecy of the sampling path, these methods achieve strong robustness against adaptive detection while preserving generation fidelity.

\vspace{-0.5mm}
\section{Evaluation Metrics for ID Technologies}
\label{sec:metrics}




An Implicit-ID method should be evaluated as a claim-verification system: given a suspect asset and a claimed source, owner, or generation process, the verifier accepts or rejects the claim under a specified access interface, threshold, and transformation regime. Since the task and notation have been introduced in Section~\ref{section:Main Tasks in Identity Technologies}, we focus here on the metrics required for comparable evaluation.

\subsection{Correctness of Identity Claims}

Correctness measures whether the verifier accepts valid claims and rejects invalid ones. We use the true positive rate (TPR) and false positive rate (FPR):
\[
    \mathrm{TPR}
    =
    \frac{\mathrm{TP}}{\mathrm{P}},
    \qquad
    \mathrm{FPR}
    =
    \frac{\mathrm{FP}}{\mathrm{N}},
\]
where \(\mathrm{P}\) is the number of positive claims and \(\mathrm{N}\) is the number of negative claims. \(\mathrm{TP}\) counts valid claims correctly accepted, while \(\mathrm{FP}\) counts invalid claims incorrectly accepted. These quantities unify common measures used in fingerprinting and watermarking studies, including fingerprint success rate, reliability, and watermark detection accuracy~\citep{xu2025copyright,yang2025watermarking,shao2025sok}.

Identity technologies should be reported at meaningful operating points, such as \(\mathrm{TPR}\) at a fixed maximum \(\mathrm{FPR}\), or through threshold-dependent curves such as ROC or precision-recall curves. For multi-source attribution, where the verifier chooses among multiple candidate owners or source models, top-\(k\) attribution accuracy and calibrated confidence should also be reported. Calibration is especially important for non-keyed fingerprinting, whose conclusions are usually probabilistic rather than cryptographic.

\subsection{Robustness to Benign Transformations}

Identity claims must remain verifiable after ordinary post-release changes. Let \(\mathcal{T}_{\mathrm{own}}\) denote benign owner-side or user-side transformations. 
Examples include dataset filtering, preprocessing, model fine-tuning, quantisation, pruning, distillation, model merging, prompt or policy updates, decoding changes, paraphrasing, translation, and summarisation. 

For a transformation \(\tau \in \mathcal{T}_{\mathrm{own}}\), evaluation should apply the transformation to the identity-bearing asset, yielding \(A'=\tau(A)\), and then measure correctness under the same declared access and threshold regime. A useful summary is benign robustness coverage:
\[
    \mathrm{RC}_{\mathrm{own}}
    =
    \frac{
    \left|
    \left\{
    \tau \in \mathcal{T}_{\mathrm{own}}
    :
    \mathrm{TPR}_{\tau} \geq C_{\mathrm{TPR}},
    \;
    \mathrm{FPR}_{\tau} \leq C_{\mathrm{FPR}}
    \right\}
    \right|
    }{
    |\mathcal{T}_{\mathrm{own}}|
    }.
\]

\subsection{Security Against Adaptive Manipulation}


A second dimension of robustness concerns adversarial security. 
Let \(\mathcal{T}_{\mathrm{adv}}\) denote transformations chosen by an adaptive adversary to remove, evade, forge, or spoof identity evidence. 
Security evaluation must specify the adversary's knowledge and capability. 
For removal or evasion attacks, the attack succeeds when a true identity claim is rejected. For forgery or spoofing attacks, the attack succeeds when a false claim is accepted. Thus, attack success can be expressed as
\[
    \mathrm{ASR}_{\mathrm{remove}}(\tau)
    =
    1-\mathrm{TPR}_{\tau},
    \qquad
    \mathrm{ASR}_{\mathrm{forge}}(\tau)
    =
    \mathrm{FPR}_{\tau}.
\]
A compact adversarial robustness coverage can be defined analogously to benign robustness:
\[
    \mathrm{RC}_{\mathrm{adv}}
    =
    \frac{
    \left|
    \left\{
    \tau \in \mathcal{T}_{\mathrm{adv}}
    :
    \mathrm{TPR}_{\tau} \geq C_{\mathrm{TPR}},
    \;
    \mathrm{FPR}_{\tau} \leq C_{\mathrm{FPR}}
    \right\}
    \right|
    }{
    |\mathcal{T}_{\mathrm{adv}}|
    }.
\]
This metric should be interpreted within the context of the specified threat model, since its value is inherently conditioned on the adversary’s capabilities.

\subsection{Utility and Deployment Cost}

Identity technologies must also preserve the primary utility of the asset and be feasible to verify. Utility impact measures degradation caused by embedding or maintaining the identity,
\[
    \Delta U
    = U_{\mathrm{orig}} - U_{\mathrm{ID}},
\]
where \(U_{\mathrm{orig}}\) is the utility of the original asset and \(U_{\mathrm{ID}}\) is the utility after identity construction. 
For generated text, \(U\) may include perplexity, semantic similarity, task accuracy, toxicity, or human preference. For models, it may include benchmark accuracy, instruction-following quality, safety behaviour, latency, or downstream task performance. For datasets, it may include data quality and downstream model performance. Watermarking usually requires careful reporting of \(\Delta U\) because it modifies the asset or generation process. Fingerprinting is non-intrusive and often has negligible utility impact, but it imposes high verification costs.

\emph{Deployment Cost} primarily refers to the computational overhead incurred by an identity technology. According to the stages at which identity is established and verified, we decompose the cost into embedding cost and verification cost.
Embedding cost is the resource required to establish an identity. For watermarking, this includes data construction, fine-tuning, model editing, logit intervention, key management, or changes to the serving pipeline. For fingerprinting, explicit embedding cost is typically zero, but reference evidence extraction and storage should still be reported. 
Verification cost includes query volume, latency, compute, memory, auxiliary storage, access requirements, and human audit effort. These costs are central to deployability: a highly identifiable method may still be impractical if it requires white-box access to a frontier-scale model, thousands of queries, or expensive pairwise comparisons across many candidate sources.

\vspace{-0.5mm}
\section{Current Challenges and Future Directions}

\paragraph{Identity signals that preserve utility.}
A persistent bottleneck is the identifiability–utility trade-off: many Implicit-ID mechanisms improve verification performance only at the cost of nontrivial utility degradation $\Delta U$ or by exploiting fragile surface-level cues. Future work should move beyond heuristic signal design and co-design identity mechanisms with \emph{semantic fidelity} as a first-class objective, achieving high separability across prompts, domains, and access interfaces while preserving model utility and user experience.

\paragraph{Robustness across lifecycle drift and adaptive transformations.}
Existing methods often fail when models undergo routine adaptation (e.g., fine-tuning, compression) or when adversaries attempt removal/forgery.
A central direction is to formalise robustness over explicit transformation suites $\mathcal{T}=\mathcal{T}_{\text{own}}\cup\mathcal{T}_{\text{adv}}$ and develop transformation-invariant identity signals and decision rules that remain valid under post-release drift, not merely in the original parameterisation.

\paragraph{Deployable verification with guarantees and shared benchmarks.}
Practical adoption is limited by verification cost (query volume, latency, compute) and by the lack of rigorous security statements.
Future work should pursue sample-efficient, auditable verification and align claims with verification semantics: non-keyed attribution as calibrated statistical evidence, keyed verification with explicit notions such as unforgeability and non-removability.
Community progress also requires standardised, cross-asset benchmarks that fix interfaces, stressors, and reporting (identifiability, robustness coverage, $\Delta U$, and verification cost) to enable credible cross-paper comparison.

\vspace{-0.5mm}
\section{Conclusions}
\label{section:Conclusions}
LLM identity technologies are becoming essential for protecting LLM-related assets and tracing their provenance. This survey has introduced \emph{Implicit-ID} as a unifying abstraction for understanding fingerprinting and watermarking, and has organised existing methods across datasets, models, and generated content under two verification semantics: similarity-based attribution and keyed verification, and highlights the metrics that matter for deployment.

LLM identity is best understood as verification under distribution shift induced by benign adaptation and adaptive attacks.
We hope this survey helps the community converge on reproducible benchmarks, consistent terminology, and practically defensible identity for trustworthy LLM deployment.

\clearpage
\section*{Acknowledgements}
\vspace{0.2em}
The authors would like to thank the anonymous reviewers and chairs for their constructive comments. This work was supported by the National Natural Science Foundation of China under Grants 62572382, 62506292, and U22A2098; the China Postdoctoral Science Foundation under Grants 2024M762604 and BX20250380; the Science and Technology Project of State Grid Headquarters under Grant 52199925001B-132-ZN; the Henan Key Laboratory of Network Cryptography Technology under Grant LNCT2025004; and the Shenzhen Loop Area Institute.

\appendix




\bibliographystyle{named}
\bibliography{ijcai26}

@inproceedings{zhang2024easydetector,
  title={EasyDetector: Using Linear Probe to Detect the Provenance of Large Language Models},
  author={Zhang, Jie and Li, Jiayuan and Fei, Haiqiang and Li, Lun and Zhu, Hongsong},
  booktitle={2024 IEEE 23rd International Conference on Trust, Security and Privacy in Computing and Communications (TrustCom)},
  pages={2410--2417},
  year={2024},
  organization={IEEE}
}

@article{wang2025building,
  title={Building intelligence identification system via large language model watermarking: a survey and beyond},
  author={Wang, Xuhong and Jiang, Haoyu and Yu, Yi and Yu, Jingru and Lin, Yilun and Yi, Ping and Wang, Yingchun and Qiao, Yu and Li, Li and Wang, Fei-Yue},
  journal={Artificial Intelligence Review},
  volume={58},
  number={8},
  pages={249},
  year={2025},
  publisher={Springer}
}

@article{liang2026watermarking,
  title={Watermarking techniques for large language models: A survey},
  author={Liang, Yuqing and Xiao, Jiancheng and Gan, Wensheng and Yu, Philip S},
  journal={Artificial Intelligence Review},
  volume={59},
  number={2},
  pages={74},
  year={2026},
  publisher={Springer}
}

@article{shao2025sok,
  title={Sok: Large language model copyright auditing via fingerprinting},
  author={Shao, Shuo and Li, Yiming and He, Yu and Yao, Hongwei and Yang, Wenyuan and Tao, Dacheng and Qin, Zhan},
  journal={arXiv preprint arXiv:2508.19843},
  year={2025}
}

@inproceedings{chen2022copy,
  title={Copy, right? a testing framework for copyright protection of deep learning models},
  author={Chen, Jialuo and Wang, Jingyi and Peng, Tinglan and Sun, Youcheng and Cheng, Peng and Ji, Shouling and Ma, Xingjun and Li, Bo and Song, Dawn},
  booktitle={2022 IEEE symposium on security and privacy (SP)},
  pages={824--841},
  year={2022},
  organization={IEEE}
}

@INPROCEEDINGS{wu2025gradient,
  author={Wu, Zehao and Zhao, Yanjie and Wang, Haoyu},
  booktitle={2025 40th IEEE/ACM International Conference on Automated Software Engineering (ASE)}, 
  title={TensorGuard: Gradient-Based Model Fingerprinting for LLM Similarity Detection and Family Classification}, 
  year={2025},
  volume={},
  number={},
  pages={445-456},
  doi={10.1109/ASE63991.2025.00044}}

@article{shao2025reading,
  title={Reading Between the Lines: Towards Reliable Black-box LLM Fingerprinting via Zeroth-order Gradient Estimation},
  author={Shao, Shuo and Li, Yiming and Yao, Hongwei and Chen, Yifei and Yang, Yuchen and Qin, Zhan},
  journal={arXiv preprint arXiv:2510.06605},
  year={2025}
}

@inproceedings{russinovich2024hey,
  title     = {{Hey, That's My Model! Introducing Chain \& Hash, an LLM Fingerprinting Technique}},
  author    = {Russinovich, Mark and Cai, Yanan and Salem, Ahmed},
  booktitle = {International Conference on Learning Representations},
  year      = {2026},
  url       = {https://mlanthology.org/iclr/2026/russinovich2026iclr-hey/}
}

@article{zeng2024huref,
  title={Huref: Human-readable fingerprint for large language models},
  author={Zeng, Boyi and Wang, Lizheng and Hu, Yuncong and Xu, Yi and Zhou, Chenghu and Wang, Xinbing and Yu, Yu and Lin, Zhouhan},
  journal={Advances in Neural Information Processing Systems},
  volume={37},
  pages={126332--126362},
  year={2024}
}

@inproceedings{pasquini2024llmmap,
title = "LLMmap: Fingerprinting for Large Language Models",
author = "Dario Pasquini and Kornaropoulos, \{Evgenios M.\} and Giuseppe Ateniese",
note = "Publisher Copyright: {\textcopyright} 2025 by The USENIX Association All Rights Reserved.; 34th USENIX Security Symposium, USENIX Security 2025 ; Conference date: 13-08-2025 Through 15-08-2025",
year = "2025",
language = "English",
series = "Proceedings of the 34th USENIX Security Symposium",
pages = "299--318",
booktitle = "Proceedings of the 34th USENIX Security Symposium",
}

@inproceedings{zhang2024reef,
 author = {Zhang, Jie and Liu, Dongrui and Qian, Chen and Zhang, Linfeng and Liu, Yong and Qiao, Yu and Shao, Jing},
 booktitle = {International Conference on Learning Representations},
 editor = {Y. Yue and A. Garg and N. Peng and F. Sha and R. Yu},
 pages = {48092--48117},
 title = {REEF: Representation Encoding Fingerprints for Large Language Models},
 url = {https://proceedings.iclr.cc/paper_files/paper/2025/file/77e830bd802f4b63b682038d090a97b6-Paper-Conference.pdf},
 volume = {2025},
 year = {2025}
}

@inproceedings{nasery2025scalable,
 author = {Nasery, Anshul and Hayase, Jonathan and Brooks, Creston and Sheng, Peiyao and Tyagi, Himanshu and Viswanath, Pramod and Oh, Sewoong},
 booktitle = {Advances in Neural Information Processing Systems},
 editor = {D. Belgrave and C. Zhang and H. Lin and R. Pascanu and P. Koniusz and M. Ghassemi and N. Chen},
 pages = {125116--125152},
 publisher = {Curran Associates, Inc.},
 title = {Scalable Fingerprinting of Large Language Models},
 url = {https://proceedings.neurips.cc/paper_files/paper/2025/file/b5ee4715ef8e96176ef3ccbe229b6ab9-Paper-Conference.pdf},
 volume = {38},
 year = {2025}
}

@InProceedings{sun2025idiosyncrasies,
  title = 	 {Idiosyncrasies in Large Language Models},
  author =       {Sun, Mingjie and Yin, Yida and Xu, Zhiqiu and Kolter, J Zico and Liu, Zhuang},
  booktitle = 	 {Proceedings of the 42nd International Conference on Machine Learning},
  pages = 	 {57854--57885},
  year = 	 {2025},
  editor = 	 {Singh, Aarti and Fazel, Maryam and Hsu, Daniel and Lacoste-Julien, Simon and Berkenkamp, Felix and Maharaj, Tegan and Wagstaff, Kiri and Zhu, Jerry},
  volume = 	 {267},
  series = 	 {Proceedings of Machine Learning Research},
  month = 	 {13--19 Jul},
  publisher =    {PMLR},
  url = 	 {https://proceedings.mlr.press/v267/sun25z.html}
}

@article{suzuki2025natural,
  title={Natural Fingerprints of Large Language Models},
  author={Suzuki, Teppei and Ri, Ryokan and Takase, Sho},
  journal={arXiv preprint arXiv:2504.14871},
  year={2025}
}

@inproceedings{antoun2023text,
    title = "From Text to Source: Results in Detecting Large Language Model-Generated Content",
    author = "Antoun, Wissam  and
      Sagot, Beno{\^i}t  and
      Seddah, Djam{\'e}",
    editor = "Calzolari, Nicoletta  and
      Kan, Min-Yen  and
      Hoste, Veronique  and
      Lenci, Alessandro  and
      Sakti, Sakriani  and
      Xue, Nianwen",
    booktitle = "Proceedings of the 2024 Joint International Conference on Computational Linguistics, Language Resources and Evaluation (LREC-COLING 2024)",
    month = may,
    year = "2024",
    address = "Torino, Italia",
    publisher = "ELRA and ICCL",
    url = "https://aclanthology.org/2024.lrec-main.665/",
    pages = "7531--7543",
}

@article{iourovitski2024hide,
  title={Hide and seek: Fingerprinting large language models with evolutionary learning},
  author={Iourovitski, Dmitri and Sharma, Sanat and Talwar, Rakshak},
  journal={arXiv preprint arXiv:2408.02871},
  year={2024}
}

@inproceedings{jin2024proflingo,
  title={Proflingo: A fingerprinting-based intellectual property protection scheme for large language models},
  author={Jin, Heng and Zhang, Chaoyu and Shi, Shanghao and Lou, Wenjing and Hou, Y Thomas},
  booktitle={2024 IEEE Conference on Communications and Network Security (CNS)},
  pages={1--9},
  year={2024},
  organization={IEEE}
}

@inproceedings{gubri2024trap,
    title = "{TRAP}: Targeted Random Adversarial Prompt Honeypot for Black-Box Identification",
    author = "Gubri, Martin  and
      Ulmer, Dennis  and
      Lee, Hwaran  and
      Yun, Sangdoo  and
      Oh, Seong Joon",
    editor = "Ku, Lun-Wei  and
      Martins, Andre  and
      Srikumar, Vivek",
    booktitle = "Findings of the Association for Computational Linguistics: ACL 2024",
    month = aug,
    year = "2024",
    address = "Bangkok, Thailand",
    publisher = "Association for Computational Linguistics",
    url = "https://aclanthology.org/2024.findings-acl.683/",
    doi = "10.18653/v1/2024.findings-acl.683",
    pages = "11496--11517"
}

@article{xu2024fp,
  title={Fp-vec: Fingerprinting large language models via efficient vector addition},
  author={Xu, Zhenhua and Xing, Wenpeng and Wang, Zhebo and Hu, Chang and Jie, Chen and Han, Meng},
  journal={arXiv preprint arXiv:2409.08846},
  year={2024}
}

@article{bitton2025detecting,
  title={Detecting Stylistic Fingerprints of Large Language Models},
  author={Bitton, Yehonatan and Bitton, Elad and Nisan, Shai},
  journal={arXiv preprint arXiv:2503.01659},
  year={2025}
}

@article{li2025editmark,
  title={EditMark: Watermarking Large Language Models based on Model Editing},
  author={Li, Shuai and Chen, Kejiang and Jiang, Jun and Zhang, Jie and Yao, Qiyi and Zeng, Kai and Zhang, Weiming and Yu, Nenghai},
  journal={arXiv preprint arXiv:2510.16367},
  year={2025}
}

@article{xu2025rap,
  title={RAP-SM: Robust Adversarial Prompt via Shadow Models for Copyright Verification of Large Language Models},
  author={Xu, Zhenhua and Wang, Zhebo and Li, Maike and Xing, Wenpeng and Hu, Chunqiang and Zhi, Chen and Han, Meng},
  journal={arXiv preprint arXiv:2505.06304},
  year={2025}
}

@article{he2025routemark,
  title={RouteMark: A Fingerprint for Intellectual Property Attribution in Routing-based Model Merging},
  author={He, Xin and Shen, Junxi and Tang, Zhenheng and Chu, Xiaowen and Li, Bo and Tsang, Ivor W and Ong, Yew-Soon},
  journal={arXiv preprint arXiv:2508.01784},
  year={2025}
}

@article{mo2025tibw,
  title={TIBW: Task-Independent Backdoor Watermarking with Fine-Tuning Resilience for Pre-Trained Language Models.},
  author={Mo, Weichuan and Chen, Kongyang and Xiao, Yatie},
  journal={Mathematics (2227-7390)},
  volume={13},
  number={2},
  year={2025}
}

@inproceedings{lin2024malla,
  title={Malla: Demystifying real-world large language model integrated malicious services},
  author={Lin, Zilong and Cui, Jian and Liao, Xiaojing and Wang, XiaoFeng},
  booktitle={33rd USENIX Security Symposium (USENIX Security 24)},
  pages={4693--4710},
  year={2024}
}

@article{xu2025copyright,
  title={Copyright Protection for Large Language Models: A Survey of Methods, Challenges, and Trends},
  author={Xu, Zhenhua and Yue, Xubin and Wang, Zhebo and Liu, Qichen and Zhao, Xixiang and Zhang, Jingxuan and Zeng, Wenjun and Xing, Wengpeng and Kong, Dezhang and Lin, Changting and others},
  journal={arXiv preprint arXiv:2508.11548},
  year={2025}
}

@article{wang2025fpedit,
  title={FPEdit: Robust LLM Fingerprinting through Localized Knowledge Editing},
  author={Wang, Shida and Liu, Chaohu and Wang, Yubo and Xu, Linli},
  journal={arXiv preprint arXiv:2508.02092},
  year={2025}
}

@article{yang2025watermarking,
  title={Watermarking for large language models: A survey},
  author={Yang, Zhiguang and Zhao, Gejian and Wu, Hanzhou},
  journal={Mathematics},
  volume={13},
  number={9},
  pages={1420},
  year={2025},
  publisher={MDPI}
}

@article{Ren2025CoTSRFUC,
  title={CoTSRF: Utilize Chain of Thought as Stealthy and Robust Fingerprint of Large Language Models},
  author={Zhenzhen Ren and Guobiao Li and Sheng Li and Zhenxing Qian and Xinpeng Zhang},
  journal={ArXiv},
  year={2025},
  volume={abs/2505.16785},
  url={https://api.semanticscholar.org/CorpusID:278788713}
}

@article{Yan2025DuFFinAD,
  title={DuFFin: A Dual-Level Fingerprinting Framework for LLMs IP Protection},
  author={Yuliang Yan and Haochun Tang and Shuo Yan and Enyan Dai},
  journal={ArXiv},
  year={2025},
  volume={abs/2505.16530},
  url={https://api.semanticscholar.org/CorpusID:278789442}
}

@article{yoon2025intrinsic,
  title={Intrinsic Fingerprint of LLMs: Continue Training is NOT All You Need to Steal A Model!},
  author={Yoon, Do-hyeon and Chun, Minsoo and Allen, Thomas and M{\"u}ller, Hans and Wang, Min and Sharma, Rajesh},
  journal={arXiv preprint arXiv:2507.03014},
  year={2025}
}

@article{zeng2025awm,
  title={AWM: Accurate Weight-Matrix Fingerprint for Large Language Models},
  author={Zeng, Boyi and Chen, Lin and He, Ziwei and Wang, Xinbing and Lin, Zhouhan},
  journal={arXiv preprint arXiv:2510.06738},
  year={2025}
}

@inproceedings{xu2024instructional,
  title={Instructional fingerprinting of large language models},
  author={Xu, Jiashu and Wang, Fei and Ma, Mingyu and Koh, Pang Wei and Xiao, Chaowei and Chen, Muhao},
  booktitle={Proceedings of the 2024 Conference of the North American Chapter of the Association for Computational Linguistics: Human Language Technologies (Volume 1: Long Papers)},
  pages={3277--3306},
  year={2024}
}

@inproceedings{kirchenbauer2023watermark,
  title={A watermark for large language models},
  author={Kirchenbauer, John and Geiping, Jonas and Wen, Yuxin and Katz, Jonathan and Miers, Ian and Goldstein, Tom},
  booktitle={International Conference on Machine Learning},
  pages={17061--17084},
  year={2023},
  organization={PMLR}
}

@inproceedings{christ2024undetectable,
  title={Undetectable watermarks for language models},
  author={Christ, Miranda and Gunn, Sam and Zamir, Or},
  booktitle={The Thirty Seventh Annual Conference on Learning Theory},
  pages={1125--1139},
  year={2024},
  organization={PMLR}
}

@inproceedings{hu2023unbiased,
 author = {Hu, Zhengmian and Chen, Lichang and Wu, Xidong and Wu, Yihan and Zhang, Hongyang and Huang, Heng},
 booktitle = {International Conference on Learning Representations},
 editor = {B. Kim and Y. Yue and S. Chaudhuri and K. Fragkiadaki and M. Khan and Y. Sun},
 pages = {45408--45436},
 title = {Unbiased Watermark for Large Language Models},
 url = {https://proceedings.iclr.cc/paper_files/paper/2024/file/c5b00c5bdcc6fe35907dbcca03d27652-Paper-Conference.pdf},
 volume = {2024},
 year = {2024}
}

@article{dathathri2024scalable,
  title={Scalable watermarking for identifying large language model outputs},
  author={Dathathri, Sumanth and See, Abigail and Ghaisas, Sumedh and Huang, Po-Sen and McAdam, Rob and Welbl, Johannes and Bachani, Vandana and Kaskasoli, Alex and Stanforth, Robert and Matejovicova, Tatiana and others},
  journal={Nature},
  volume={634},
  number={8035},
  pages={818--823},
  year={2024},
  publisher={Nature Publishing Group UK London}
}

@inproceedings{hou2024semstamp,
  title={Semstamp: A semantic watermark with paraphrastic robustness for text generation},
  author={Hou, Abe and Zhang, Jingyu and He, Tianxing and Wang, Yichen and Chuang, Yung-Sung and Wang, Hongwei and Shen, Lingfeng and Van Durme, Benjamin and Khashabi, Daniel and Tsvetkov, Yulia},
  booktitle={Proceedings of the 2024 Conference of the North American Chapter of the Association for Computational Linguistics: Human Language Technologies (Volume 1: Long Papers)},
  pages={4067--4082},
  year={2024}
}

@article{xu2025mark,
  title={Mark your llm: Detecting the misuse of open-source large language models via watermarking},
  author={Xu, Yijie and Liu, Aiwei and Hu, Xuming and Wen, Lijie and Xiong, Hui},
  journal={arXiv preprint arXiv:2503.04636},
  year={2025}
}

@inproceedings{xu2024learning,
  title     = {{Learning to Watermark LLM-Generated Text via Reinforcement Learning}},
  author    = {Xu, Xiaojun and Yao, Yuanshun and Liu, Yang},
  booktitle = {ICLR 2025 Workshops: WMARK},
  year      = {2025},
  url       = {https://mlanthology.org/iclrw/2025/xu2025iclrw-learning/}
}

@inproceedings{hou2024k,
  title={k-SemStamp: A clustering-based semantic watermark for detection of machine-generated text},
  author={Hou, Abe and Zhang, Jingyu and Wang, Yichen and Khashabi, Daniel and He, Tianxing},
  booktitle={Findings of the Association for Computational Linguistics: ACL 2024},
  pages={1706--1715},
  year={2024}
}

@inproceedings{gu2023learnability,
 author = {Gu, Chenchen and LI, XIANG and Liang, Percy and Hashimoto, Tatsunori},
 booktitle = {International Conference on Learning Representations},
 editor = {B. Kim and Y. Yue and S. Chaudhuri and K. Fragkiadaki and M. Khan and Y. Sun},
 pages = {38745--38768},
 title = {On the Learnability of Watermarks for Language Models},
 url = {https://proceedings.iclr.cc/paper_files/paper/2024/file/a86d17b6cd70366d56ab48d2a05a4df1-Paper-Conference.pdf},
 volume = {2024},
 year = {2024}
}

@article{ye2025securing,
  title={Securing Large Language Models: A Survey of Watermarking and Fingerprinting Techniques},
  author={Ye, Peigen and Ren, Huali and Li, Zhengdao and Yan, Anli and Yan, Hongyang and Wang, Shaowei and Li, Jin},
  journal={ACM Computing Surveys},
  year={2025},
  publisher={ACM New York, NY}
}

@inproceedings{maini2021dataset,
  author       = {Pratyush Maini and Mohammad Yaghini and Nicolas Papernot},
  title        = {Dataset Inference: Ownership Resolution in Machine Learning},
  booktitle    = {International Conference on Learning Representations},
  year         = {2021}
}

@article{liu2022your,
  title={Your model trains on my data? Protecting intellectual property of training data via membership fingerprint authentication},
  author={Liu, Gaoyang and Xu, Tianlong and Ma, Xiaoqiang and Wang, Chen},
  journal={IEEE Transactions on Information Forensics and Security},
  volume={17},
  pages={1024--1037},
  year={2022},
  publisher={IEEE}
}

@inproceedings{du2025sok,
  title={Sok: Dataset copyright auditing in machine learning systems},
  author={Du, Linkang and Zhou, Xuanru and Chen, Min and Zhang, Chusong and Su, Zhou and Cheng, Peng and Chen, Jiming and Zhang, Zhikun},
  booktitle={2025 IEEE Symposium on Security and Privacy (SP)},
  pages={1--19},
  year={2025},
  organization={IEEE}
}

@article{maini2024llm,
  title={LLM Dataset Inference: Did you train on my dataset?},
  author={Maini, Pratyush and Jia, Hengrui and Papernot, Nicolas and Dziedzic, Adam},
  journal={Advances in Neural Information Processing Systems},
  volume={37},
  pages={124069--124092},
  year={2024}
}

@inproceedings{carlini2021extracting,
  title={Extracting training data from large language models},
  author={Carlini, Nicholas and Tramer, Florian and Wallace, Eric and Jagielski, Matthew and Herbert-Voss, Ariel and Lee, Katherine and Roberts, Adam and Brown, Tom and Song, Dawn and Erlingsson, Ulfar and others},
  booktitle={30th USENIX security symposium (USENIX Security 21)},
  pages={2633--2650},
  year={2021}
}

@inproceedings{shi2023detecting,
 author = {Shi, Weijia and Ajith, Anirudh and Xia, Mengzhou and Huang, Yangsibo and Liu, Daogao and Blevins, Terra and Chen, Danqi and Zettlemoyer, Luke},
 booktitle = {International Conference on Learning Representations},
 editor = {B. Kim and Y. Yue and S. Chaudhuri and K. Fragkiadaki and M. Khan and Y. Sun},
 pages = {51826--51843},
 title = {Detecting Pretraining Data from Large Language Models},
 url = {https://proceedings.iclr.cc/paper_files/paper/2024/file/e32ad85fa27be4a9868d55703f01323e-Paper-Conference.pdf},
 volume = {2024},
 year = {2024}
}

@inproceedings{mattern2023membership,
    title = "Membership Inference Attacks against Language Models via Neighbourhood Comparison",
    author = {Mattern, Justus  and
      Mireshghallah, Fatemehsadat  and
      Jin, Zhijing  and
      Sch{\"o}lkopf, Bernhard  and
      Sachan, Mrinmaya  and
      Berg-Kirkpatrick, Taylor},
    editor = "Rogers, Anna  and
      Boyd-Graber, Jordan  and
      Okazaki, Naoaki",
    booktitle = "Findings of the Association for Computational Linguistics: ACL 2023",
    month = jul,
    year = "2023",
    address = "Toronto, Canada",
    publisher = "Association for Computational Linguistics",
    url = "https://aclanthology.org/2023.findings-acl.719/",
    doi = "10.18653/v1/2023.findings-acl.719",
    pages = "11330--11343"
}

@article{li2025data,
  title={Data Provenance Auditing of Fine-Tuned Large Language Models with a Text-Preserving Technique},
  author={Li, Yanming and Ghozzi, Seifeddine and Eichler, C{\'e}dric and Anciaux, Nicolas and Bensamoun, Alexandra and Manzano, Lorena Gonzalez},
  journal={arXiv preprint arXiv:2510.09655},
  year={2025}
}

@article{naser2025auditing,
  title={Auditing the Shadows: A Review of Methods to Detect Shared Training Data in Large Language Models},
  author={Naser, MZ},
  journal={ACM Computing Surveys},
  volume={58},
  number={7},
  pages={1--34},
  year={2025},
  publisher={ACM New York, NY}
}

@inproceedings{huang2024general,
  title={A general framework for data-use auditing of ML models},
  author={Huang, Zonghao and Gong, Neil Zhenqiang and Reiter, Michael K},
  booktitle={Proceedings of the 2024 on ACM SIGSAC Conference on Computer and Communications Security},
  pages={1300--1314},
  year={2024}
}

@article{liu2024survey,
  title={A survey of text watermarking in the era of large language models},
  author={Liu, Aiwei and Pan, Leyi and Lu, Yijian and Li, Jingjing and Hu, Xuming and Zhang, Xi and Wen, Lijie and King, Irwin and Xiong, Hui and Yu, Philip},
  journal={ACM Computing Surveys},
  volume={57},
  number={2},
  pages={1--36},
  year={2024},
  publisher={ACM New York, NY}
}

@article{sadhukhan2025footprints,
  title={Footprints of Data in a Classifier: Understanding the Privacy Risks and Solution Strategies},
  author={Sadhukhan, Payel and Chakraborty, Tanujit},
  journal={Information Systems Frontiers},
  pages={1--27},
  year={2025},
  publisher={Springer}
}

@article{hou2025fixguard,
  title={FixGuard: Repairing Backdoored Models via Class-Wise Trigger Recovery and Unlearning},
  author={Hou, Linshan and Hua, Zhongyun and Luo, Wei and Zhang, Leo Yu},
  journal={IEEE Signal Processing Letters},
  volume={32},
  pages={2544--2548},
  year={2025},
  doi={10.1109/LSP.2025.3572409}
}








\end{document}